\documentclass[acmtog,nonacm]{acmart}

\usepackage[utf8]{inputenc}
\usepackage{csquotes}
\usepackage[english]{babel}
\usepackage[draft]{pdfcomment}
\usepackage{subfig}
\usepackage{tikz}
\usepackage{color}
\usepackage[binary-units]{siunitx}
\usepackage[linesnumbered,ruled,vlined]{algorithm2e}

\usetikzlibrary{positioning,shapes.geometric,arrows,fit,backgrounds}

\DeclareUnicodeCharacter{202F}{\,}

\acmJournal{TOG}
\begin{document}

\title{Path Verification for Dynamic Indirect Illumination}

\author{Pierre Moreau}
\affiliation{%
  \institution{Lund University}%
  \country{Sweden}
}

\author{Michael Doggett}
\affiliation{%
  \institution{Lund University}%
  \country{Sweden}
}

\author{Erik Sintorn}
\affiliation{%
  \institution{Chalmers University of Technology}%
  \country{Sweden}
}

\begin{abstract}
In this paper we present a technique that improves rendering performance for
real-time scenes with ray traced lighting in the presence of dynamic lights and objects.
In particular we verify photon paths from the previous frame against dynamic
objects in the current frame, and show how most photon paths are still valid.
When using area lights, we use a data structure to store light distribution
that tracks light paths allowing photons to be reused when the light source is
moving in the scene.
We also show that by reusing paths when the error in the reflected energy is
below a threshold value, even more paths can be reused.
We apply this technique to Indirect Illumination using a screen space photon
splatting rendering engine.
By reusing photon paths and applying our error threshold, our method can reduce
the number of rays traced by up to 5$\times$, and improve performance by up to
2$\times$.
\end{abstract}



\begin{teaserfigure}
  \centering
  \begin{tabular}{cc}
  \includegraphics[width=8cm]{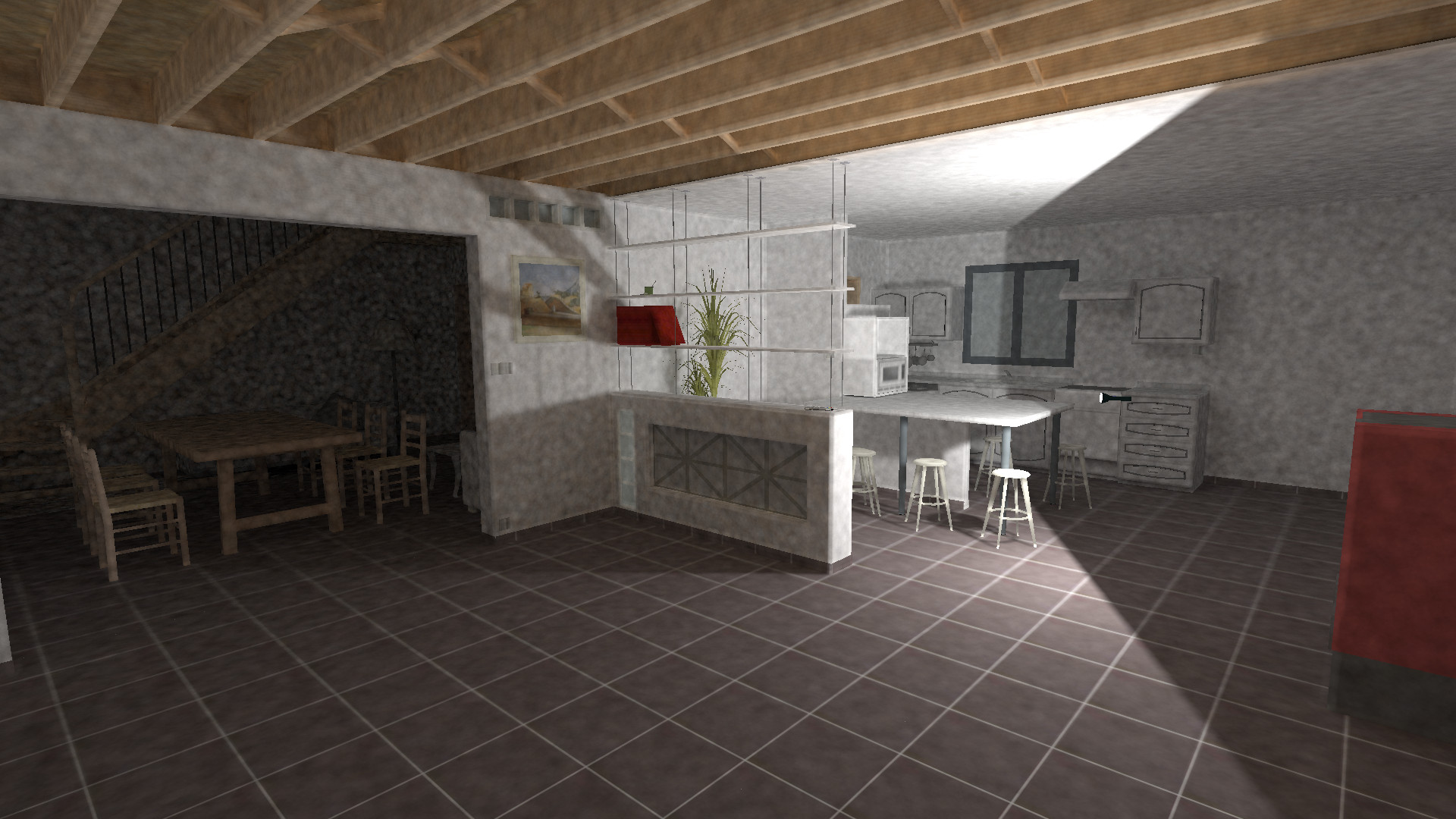} &
  \includegraphics[width=8cm]{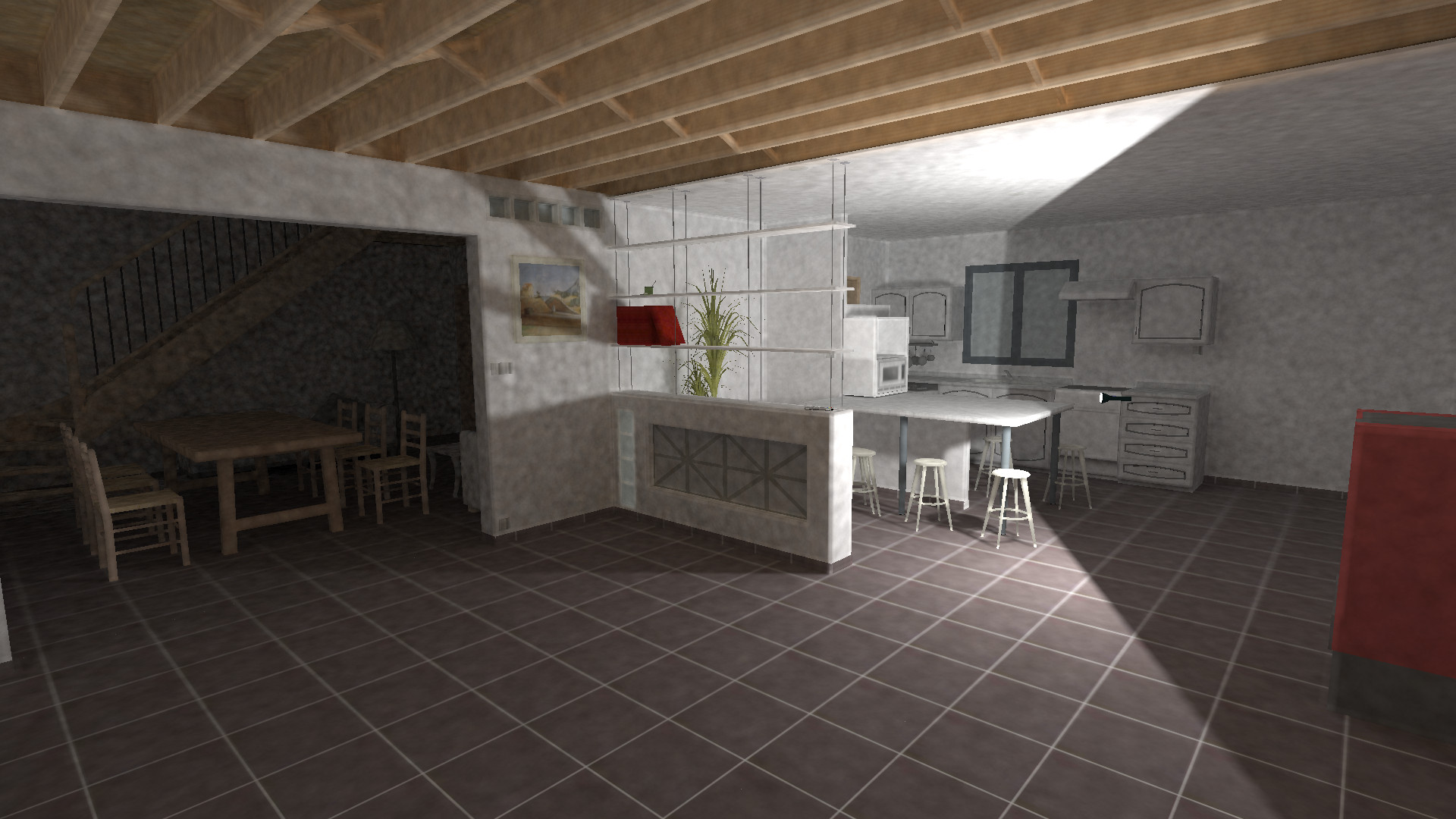}
  \end{tabular}
  \caption{Rendering in the Villa scene for the baseline on the left, and our
    error-based method on the right at roughly equal frame time (341 ms, and
    resp. 333 ms); in both cases, the number of bounces is limited to 7. The
    baseline traces 3 million paths in about 178 ms, and the splatting of the
    photons takes 148 ms. On the other hand, our method traces and reuses 5
    million paths in about 80 ms, and the splatting of the photon takes 236 ms.
  }
\end{teaserfigure}

\maketitle

\footnotetextcopyrightpermission{\copyright\ 2021\ Copyright held by the authors, published under Creative Commons\\ CC-BY-SA-4.0 License.}

\section{Introduction}

Indirect illumination is an important cue for the perceived realism of computer
generated imagery, but its accurate computation can be computationally expensive.
A recent survey by Ritschel et al.~\shortcite{ritschel12} covers many algorithms that approximate indirect illumination for real-time applications.
Since the general problem is complex, and cannot be easily solved even in offline rendering, where
hundreds of cores can spend hours on a single frame, most real-time algorithms
are specifically designed to generate a good estimation under very specific
assumptions about lighting and materials.

Computing indirect illumination and soft shadows, while considering animated objects, leads to a more accurate representation of the lighting in a scene, than can usually be achieved with pre-computed techniques.
In this paper we build upon the photon splatting technique by Moreau et al.~\shortcite{moreau16} to enable indirect lighting with multiple dynamic light sources.
Unlike Moreau et al.~\shortcite{moreau16}, where the photon map is recomputed every frame, we propose to opportunistically reuse as many photon paths as possible, including those from moving light sources.
We achieve this by reusing lighting from previous frames, and within area lights.
In this paper we present a technique that uses a photon map with several bounces, many more than previous techniques, enabling subtle lighting effects in neighboring areas which have no direct path to the light source.

Photon mapping traces light paths from the light, and deposits photon energy onto diffuse surfaces.
To create the final image, camera rays are traced to gather the light deposited on surfaces in the scene.
Path tracing instead traces rays from the camera through the scene until they find a light source. This means that for traditional path tracing, all segments of a path must be verified if lights or objects move in the scene. While for photon mapping, only the light paths from the light to the surface need to be verified, as camera rays will be traced regardless.

In scenes with moving objects and dynamic light sources, we present techniques for path verification.
If a path from a previous frame is classified as still valid, we reuse the photon path in the current frame.
By reusing photon paths we are able to achieve interactive frame rates in scenes with indirect illumination.
Unlike previous methods with geometric approximations or sparse samples sets, we use a dense photon map, and instead of recomputing all light per frame, including lighting that is still valid, we carefully reuse the light transport from the previous frame.

Using photon maps allows indirect illumination to be computed in world space without the limitations of screen space methods, and enables the possibility to temporally reuse light transport computations from previous frames.

\section{Related Work}

Early attempts to achieve precomputed light transport stored it in textures~\cite{mcTaggart04},
or used spherical harmonics to store precomputed light transport~\cite{sloan02},
or, more recently, in precomputed light field probes~\cite{mcGuire17}.
These algorithms can be very
efficient to query, but require significant computational resources for offline precomputations,
and large memory buffers for high quality results.
More importantly, they do not allow for dynamic lights. On the other end of the
spectra are screen space reflection algorithms~\cite{sousa11}
that can very quickly estimate the radiance reflected from a glossy material, but only when the
reflected surfaces are directly visible to the user.
Other screen space algorithms~\cite{ritschel09} are only effective for local reflections
when the material is lambertian and do not take into account more complex
materials or light sources outside of the viewing volume.

Dmitriev et al.~\shortcite{Dmitriev2002} used two different types of photons to detect areas where lighting changed between frames, and focused updates to the lighting in those regions using \emph{corrective photons}.
This allows them to support dynamic scenes and prioritise the updates to perform, however each corrective photons needs to be traced twice, with one of the tracings taking place against an earlier version of the scene.

To reduce the number of paths needed per pixel, Bekaert et al.~\shortcite{bekaert02} proposed creating path segments in neighboring pixels and sharing those paths to increase the number of paths traced in each individual pixel. While our work also attempts to reduce computation per frame by reusing paths, we reuse the photon paths from the previous frame, not neighboring pixels.

Voxel Cone Tracing~\cite{crassin11} alleviates both of these
problems by voxelizing a rough representation of the scene around the user's
position, and can be used for diffuse indirect illumination, but is much too
expensive in terms of memory to allow for large scenes.
Also this algorithm does not allow for many bounces of light.

Also dynamic scenes have been mixed with Stochastic Progressive Photon Mapping~\cite{weiss12}, but this is a much more complex technique
and requires much longer frame times than our technique.

More recently, several denoising algorithms have been suggested, that allow for
fast denoising of extremely noisy path-traced images~\cite{chaitanya17,schied17,mara17}. By filtering samples both
spatially and temporally, the reflected radiance of each pixel can be estimated
almost as well as if hundreds of indirect illumination rays had been shot per
pixel and can handle both glossy and diffuse surfaces.
While these approaches can generate path traced images at real-time rates, they use short path lengths to ensure performance.

Another recent method is that of Silvennoinen et al.~\shortcite{silvennoinen17} where a very sparse set of light probes are updated every frame by shooting a single ray per direction and looking up the intersected
surface's direct-lighting response in a texture.
Our technique robustly checks all dynamic objects, unlike Silvennoinen et
al.~\shortcite{silvennoinen17}, which only has support for approximated dynamic
objects and so will fail to correctly capture the illumination when all light
has undergone several bounces before reaching the camera.

Corso et al.~\shortcite{Corso2017} recently looked into reusing shading information at primary view samples from previous frames.
This is done by reprojecting the view sample locations into the current frame and validating their visibility.
They also maintain a uniform distribution of outgoing rays to avoid having too many or too few paths at a given pixel.
Our approach extends the validation to consider the whole path rather than just the first segment, and we modified how the uniform distribution is maintained to apply to light sources and support area lights.

\section{Algorithm}
\label{sec:algorithm}

\begin{figure}
  \centering
  \includegraphics[width=0.49\textwidth]{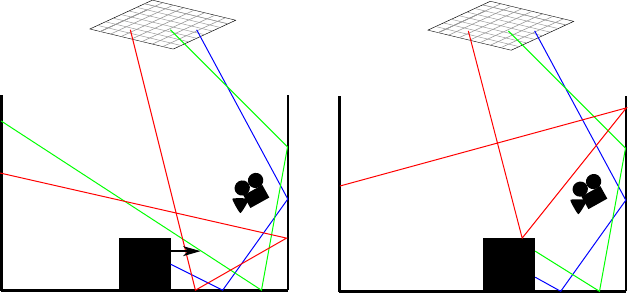}
  \caption{A simple scene showing an area light source at the top, show as a
    grid, from which several photon paths are traced. A small cube moves in the
    bottom of the scene, shown in it's starting position on the left, and
    it's new position on the right.
    In the right image the cube intersects the existing segments of all
    paths.
    The grid at the top represents what we call a distribution map covering the
    two dimensional area light.
    For the green and blue paths, as only their last segment is being
    intersected, we only need to compute the new intersection point. On the
    other hand, the red path’s first segment is intersected also triggering a
    computation of the new intersection point, but also a resampling of the BRDF
    there, due to the second photon having become invisible from the first one.
    However, since the new second photon can see the old third one, and the
    energy between the two is similar to before, the third photon is kept as-is.
  }\label{overview}
\end{figure}

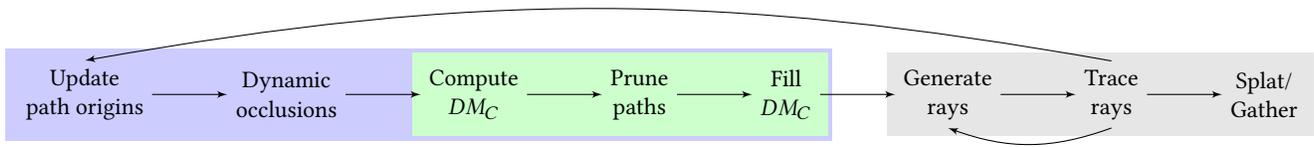
\begin{figure*}
  \centering
  \tikzstyle{arrow} = [ draw, -latex' ]
\begin{tikzpicture}[
  align=center,
  reuse block/.style = {
    fill = blue!20,
    inner sep = +0.5em
  },
  dm block/.style = {
    fill = green!20
  },
  original block/.style = {
    fill = gray!20
  }
  ]

  \node (updateOrigin)  {Update\\path origins};
  \node (dynOccl)  [right=of updateOrigin] {Dynamic\\occlusions};
  \node (setupDMs) [right=of dynOccl]  {Compute\\$DM_C$};
  \node (rmExtra)  [right=of setupDMs] {Prune\\paths};
  \node (fillCells)  [right=of rmExtra] {Fill\\$DM_C$};
  \node (genRays)  [right=of fillCells]   {Generate\\rays};
  \node (trace)    [right=of genRays]   {Trace\\rays};
  \node (render)   [right=of trace]    {Splat/\\Gather};

  \path [arrow]  (updateOrigin) -- (dynOccl);
  \path [arrow]  (dynOccl.east) -- (setupDMs.west);
  \path [arrow]      (setupDMs) -- (rmExtra);
  \path [arrow]       (rmExtra) -- (fillCells);
  \path [arrow]       (fillCells) -- (genRays.west);
  \path [arrow]       (genRays) -- (trace);
  \path [arrow]         (trace.south) to [curve to,bend left=20] (genRays.south);
  \path [arrow]         (trace.north) to [curve to,bend right=10] (updateOrigin.north);
  \path [arrow]         (trace) -- (render);

  \begin{scope}[on background layer]
    \node [reuse block,    fit=(setupDMs) (rmExtra) (fillCells) (dynOccl) (updateOrigin)] {};
    \node [dm block,       fit=(setupDMs) (rmExtra) (fillCells)] {};
    \node [original block, fit=(genRays) (trace) (render)] {};
  \end{scope}
\end{tikzpicture}
  \caption{A diagram showing where our algorithm sits in a regular photon
    renderer, visualised as a grey box, as well as its different main steps.
    The green box groups steps involving distribution maps together, whereas the
    blue box groups all steps needed for reusing photons from frame to frame.
    The \enquote{generate rays} and \enquote{trace rays} steps bare a few
    differences in our algorithm, compared to the classic version. However, as
    those are not significant, they are represented as the same steps in this
    diagram; the differences will be presented in Section~\ref{sec:implementation}.
    $DM_C$ represents the \emph{distribution map} of the current frame,
    computed from the existing light paths.
  }
  \label{fig:algorithm_diagram}
\end{figure*}

Given a scene made only of static objects and lights, the tracing of the light
paths only needs to be done once and can be reused for all frames.
In this paper we focus on the reuse of light paths from previous frames in the
presence of dynamic objects and lights, and are not concerned with static
scenes.
To verify that a light path is still valid in the current frame, it must be
checked against moving objects and light sources.
Figure~\ref{overview} illustrates how the algorithm handles a single moving
object intersecting three photon paths.
In this section we outline how this verification of light paths is performed
first for dynamic lights, and then for dynamic objects.

Our algorithm is made of 5 main steps, that process all light paths from
previous frames and tell a slightly modified photon mapper/splatter which paths
should be retraced; details about the modifications done to the photon
mapper/splatter can be found in Section~\ref{sec:implementation}. In the
following we give a brief introduction to the five main stages of our algorithm,
and then explain them in more detail later in this section.
\begin{description}
  \item[Update path origins]  to match the current position and orientation of
    light sources.
  \item[Dynamic occlusions] will detect and schedule for re-tracing paths
    intersected by dynamic objects.
  \item[Compute $DM_C$] to know how many light paths are emitted from each cell.
  \item[Prune paths] to decrease the number of emitted paths for cells with too
    many light paths.
  \item[Fill $DM_C$] to increase the number of light paths in cells below the
    required amount.
\end{description}
Apart from \enquote{dynamic occlusions}, which will be presented in
Section~\ref{sec:dynObjects} as it is unnecessary for dynamic lights, the
remaining four main steps will be presented in Section~\ref{sec:dynLights}.
Figure~\ref{fig:algorithm_diagram} shows the structure of the five steps
and where they sit in relation to a regular photon splatting architecture.

\subsection{Supporting dynamic lights}\label{sec:dynLights}

\begin{figure}
  \centering
  \includegraphics[width=0.4\textwidth]{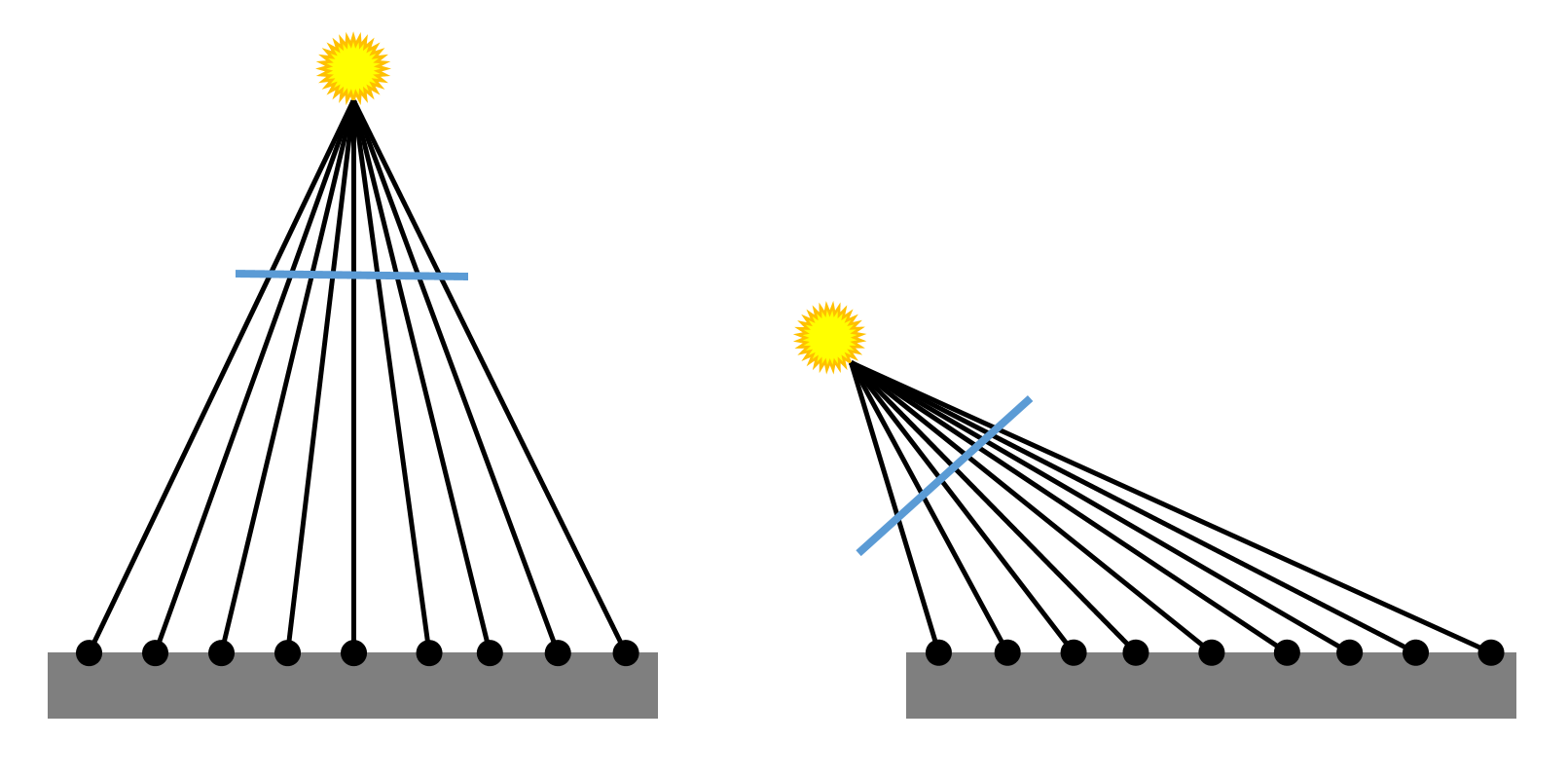}
  \caption{On the left, a spotlight is lighting a certain area on the
    ground, with the primary photons represented as black spheres and its near
    plane as a blue line. If one was to keep the same photons as the light
    moves (while still illuminating the same area), it would result in a
    distribution of light across the near plane that is different from the
    initial one, as seen on the right.
  }\label{fig:different_angle_lighting}
\end{figure}

As lights move in a scene, some surfaces that were previously unlit become
now visible to a light source and receive light, while others fade in the
shadows. To validate light paths against such behaviours, a first approach would
be to test whether the primary segment of each path is still within the light's
field of view, and if not, replace the whole path by a new one. However this can
lead to changes in the light’s distribution, as showcased in
Figure~\ref{fig:different_angle_lighting}.

Even a light that moves parallel to the plane it is illuminating, will have
issues if photons that are now no longer visible from the light are randomly
re-traced over the whole volume visible from the light. This would result in
very few photons in the newly visible areas, as many of the new photons would
end up in the already visible areas. To avoid those issues, we propose to
maintain the distribution of photons from the light source between frames.

\begin{figure}
  \centering
  \includegraphics[width=0.4\textwidth]{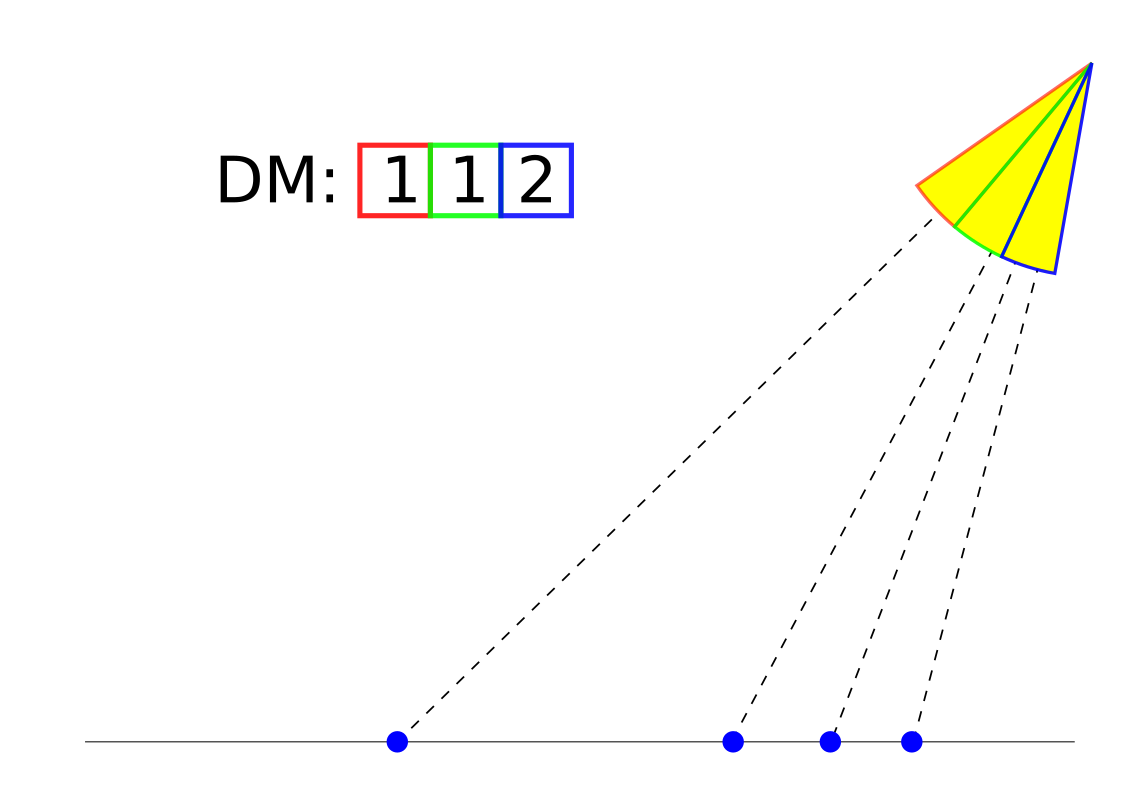}
  \caption{A simple example of a distribution map for a spotlight in 2D. Here
    the distribution map is composed of only three cells, and is represented in
    the top-left corner as an array. Each cell keeps track of how many paths
    originated from a specific region on the light; the cells form a partition
    of all possible origin configurations. The mapping between a cell and its
    corresponding region on the light is colour-coded and can be visualised
    directly on the figure.
  }\label{fig:simple_dm_example}
\end{figure}

To achieve equal distribution across the light we partition its surface, and its
set of outgoing directions into cells, and ensure that each cell maintains
a given amount of primary paths emitted from that cell. Those cells form an
$n$-D array which we call a \emph{Distribution Map} (DM). The parametrisation of
this array is not constrained and can be different for different light types.
For example, a point light could have a 2-D parametrisation $(\theta,\phi)$ whereas a
rectangular area light could use a 4-D parametrisation $(x,y,\theta,\phi)$; an
example of parametrisation for a 2-D spotlight can be seen in
Figure~\ref{fig:simple_dm_example}.

The distribution map is initialised with a user-defined distribution for the
light source. This initial set of values is denoted as $DM_T$, or the
\emph{targeted} distribution of the light. This target distribution could be
updated every frame to allow for textured light sources. A second distribution
map, noted $DM_C$, is computed each frame using all existing paths at the end of
the previous frame. Each frame, we apply a set of operations to make $DM_C$
converge towards $DM_T$; those operations might update the values stored in
$DM_C$ to ensure that it counts only valid paths.

\paragraph{Update path origins} As the lights move, we need to update the
position on the light from which the paths are emitted. For a point light, this
simply means setting the light sample to the new position of the light and
recomputing the outgoing direction based on this new position and the existing
primary photon. We also check that the primary photon is still visible from the
light, by making use of the attached shadow map. This will however not work for
area lights, so in those cases, to avoid tracing visibility rays towards each
primary photon, we keep the existing outgoing direction and compute its
intersection with the plane of the light to get the new origin of the path. Some
of the light paths can already be invalidated during this step.

\paragraph{Compute $DM_C$} As the light path origins have been updated to
reflect the current position and orientation of the lights, we can now compute
how many light paths are emitted from each cell; this is done for all paths that
were successfully updated in the previous step. If the parametrisation function
of the distribution map returns a correct value given the position on the light
and outgoing direction of a light path, the cell found to have emitted this path
is atomically increased. Otherwise, the path is marked as invalid and will be
re-traced in the “fill $DM_C$” step.

\paragraph{Prune paths} Thanks to the previous step, we now know how many paths
lie in each cell. Some of them might contain more paths than they should, if the
light moved. In order to converge back to $DM_T$, for each cell where $DM_C >
DM_T$, every path emitted from that cell will be pruned with the following
probability:
\begin{equation} \label{eq:pruneProba}
  \frac{DM_C - DM_T}{DM_C}
\end{equation}
note that this does not ensure that $DM_C$ will be equal to $DM_T$, but ensures
that $DM_C$ will converge towards the target over a number of frames. Also, all
paths pruned by this pass are valid paths: we could keep them and reduce their
energies, however that could result over time in paths with low energy, so we
prune them instead.

\paragraph{Fill $DM_C$} For the same reasons that some cells will contain more
paths, others will be lacking some paths.
For each cell to reach its expected amount of paths, we sample the light to
obtain a new position on the light and outgoing direction.
The sampling of the light is restricted to the domain contained within the cell.
Those inputs will later be used to trace new light paths in the \enquote{trace
  rays} step.

\subsection{Supporting dynamic objects}\label{sec:dynObjects}

To handle dynamic objects, the “dynamic occlusions” step of the algorithm adds
visibility rays to compute whether the visibility between two vertices of the
path changed. These rays test the current segment against the bounding
box of every dynamic object in the scene, and kill the segment if any of the
tests fail. This is a conservative approach and might return false positives.

In order to avoid unnecessary tests, we test the segments of a path in order,
starting from the segment leaving the light. If the $i$th segment of the
path is intersected by a dynamic object, all segments after it will be
different. After a segment is found to be intersected by a dynamic object, we
schedule that segment and all the following ones to be re-traced.

\subsection{Error-based threshold for path reuse}\label{sec:errorBased}

While the solution presented in Section~\ref{sec:dynObjects} is straightforward,
some paths propagate very similar energy from frame to frame and could be
reused. Instead of killing the intersected segment $i$, we trace a visibility ray
from the segment’s origin to it’s destination, and compute the new end of that
segment, which is also the origin of the next segment $j$. As we updated the origin
of $j$, we may break the visibility between both ends of that segment. We can
trace a new visibility ray along $j$ to compute its new end point, and
continue similarly until we reach the end of the path.
The error-based threshold algorithm is shown in Algorithm~\ref{alg:dynOccl}.

We can however avoid the visibility ray under certain circumstances: if the
segment is not intersected by any dynamic objects, neither its origin nor end
are located on dynamic objects, and its newly computed origin is located at the
same position as its old origin. This situation can occur when segment $i$ is
intersected by the bounding box of a dynamic object, but in practice is not
intersected by any of the object’s triangles.

When reusing path vertices, sometimes the energy reflected $E_N$, at the new
intersection point is very similar to how much energy $E_O$, was being reflected at
the old intersection point. We detect these cases by using a user specified
energy threshold $T$, and if Equation~\ref{eq:energyThreshold} is satisfied, we
don't update or propagate the new energy value saving valuable computation
time. Otherwise, all the segments — starting from the current vertex — are
re-traced. By always comparing $E_N$ to the original reflected energy $E_O$, we
ensure that we do not accumulate errors for the energy over multiple frames.
\begin{equation}\label{eq:energyThreshold}
  (-T \times E_O \leq E_N - E_O) \land (E_N - E_O \leq T \times E_O)
\end{equation}

This technique will never trace more rays than if the whole path had been
invalidated, and can improve the temporal coherency by reusing some of the
segments.

\begin{algorithm}
  \DontPrintSemicolon
  \SetKw{Continue}{continue}\SetKw{Not}{not}\SetKw{Return}{return}
  \tcp{For each path, iterate over its segments, starting from the first one.}
  \ForEach{$Segment \in Path$}{
    \If{$\Not IntersectedByObjects(Segment)$}{\Continue}
    \tcp{Compute intersection along segment}
    $Hit \leftarrow TraceRay(Segment.origin, Segment.dir)$\;
    $HitPos \leftarrow ComputeHitPos(Segment, Hit)$\;
    $NextSeg \leftarrow Next(Segment,Path)$\;
    \tcp{Compute new outgoing direction}
    $NextSeg.origin \leftarrow HitPos$\;
    $NextSeg.dir \leftarrow NextSeg.dest - NextSeg.origin$\;
    $Energy \leftarrow BRDF(Segment,NextSeg)$\;
    \uIf{$\Not AreEnergiesClose(NextSeg.energy,Energy)$}{
      $Segment.dest \leftarrow HitPos$\;
      \tcp{Sample BRDF to generate new ray}
      \Return
    }
    \uElseIf{$AreClose(HitPos, Segment.dest) \land \Not IntersectedByObjects(NextSeg) \land
         \Not HitMovingObject(NextSeg.origin) \land \Not HitMovingObject(NextSeg.dest)$}{
      \tcp{Skip visibility check between NextSeg.origin and NextSeg.dest}
      $Segment \leftarrow NextSeg$\;
      \Continue
    }
    $Segment.dest \leftarrow HitPos$\;
    \tcp{Visibility check for NextSeg will occur on the next iteration}
  }
  \caption{Error-based threshold approach to dynamic occlusions handling}\label{alg:dynOccl}
\end{algorithm}

\section{Implementation details}\label{sec:implementation}

Each path is made up of several photons, and in order to keep track of the
paths’ structure, including the photon order, we store the photons in a 2-D
array where the $i$-th row contains the $i$-th photon of a path, and each column
is a different path. This memory layout, rather than its transpose, allows for
better memory access patterns, as all threads loading their $i$-th photon will
result in consecutive memory accesses. Photons use a total of 32 bytes:
\begin{itemize}
  \item Incoming direction (as XYZ): 3 floats;
  \item ID of object hit: 32-bit integer;
  \item Energy (as RGB): 3 floats;
  \item Radius: float.
\end{itemize}

To help with the current status of a path, we store separately a small data
structure (a single 32-bit word) containing the following: 
\begin{itemize}
  \item The ID of the $DM_C$’s cell in which this path lies. (22 bits);
  \item The number of segments in the path. (4 bits);
  \item Starting segment to retrace path from. (4 bits);
  \item Replace path. The path is retraced if the bit is set. (1 bit);
  \item Reuse light keeping light position and direction. (1 bit).
\end{itemize}

The representation of the different steps as seen in
Figure~\ref{fig:algorithm_diagram}
does not match 1:1 to our implementation. For example, we actually
update the path origins and compute the $DM_C$ in the same kernel, while the
dynamic occlusions are tested right after that. The merging of the two kernels
was done for performance reasons, in order to avoid reading from memory data
that was recently written, and the two kernels were relatively small. Since the
dynamic occlusions testing can not be done before the path origins are updated,
it had to be moved after the computation of the $DM_C$.

For simplicity reasons, we generate new rays as soon as it has been decided we
need to replace an existing ray. This means that ray generation is effectively
done in multiple places: during the dynamic occlusions testing, when filling the
$DM_C$ and when processing the results from the tracing pass, if the maximum
depth has not been reached yet.

Finally, when pruning extra paths, we end up modifying the number of paths found
in the distribution map, while needing to use the initial amount in the pruning
probability (see Equation~\eqref{eq:pruneProba}). This can be achieved by
modifying a copy of the distribution, thus using more memory, or by doing the
update in two passes by first marking the pruned paths, and then editing the
distribution map values. We are using the second approach in our implementation.

\section{Results}

\begin{figure*}
  \centering
  \subfloat[Merry-go-round]{\label{fig:confMovingBunniesTeapots}
    \includegraphics[width=0.32\textwidth]{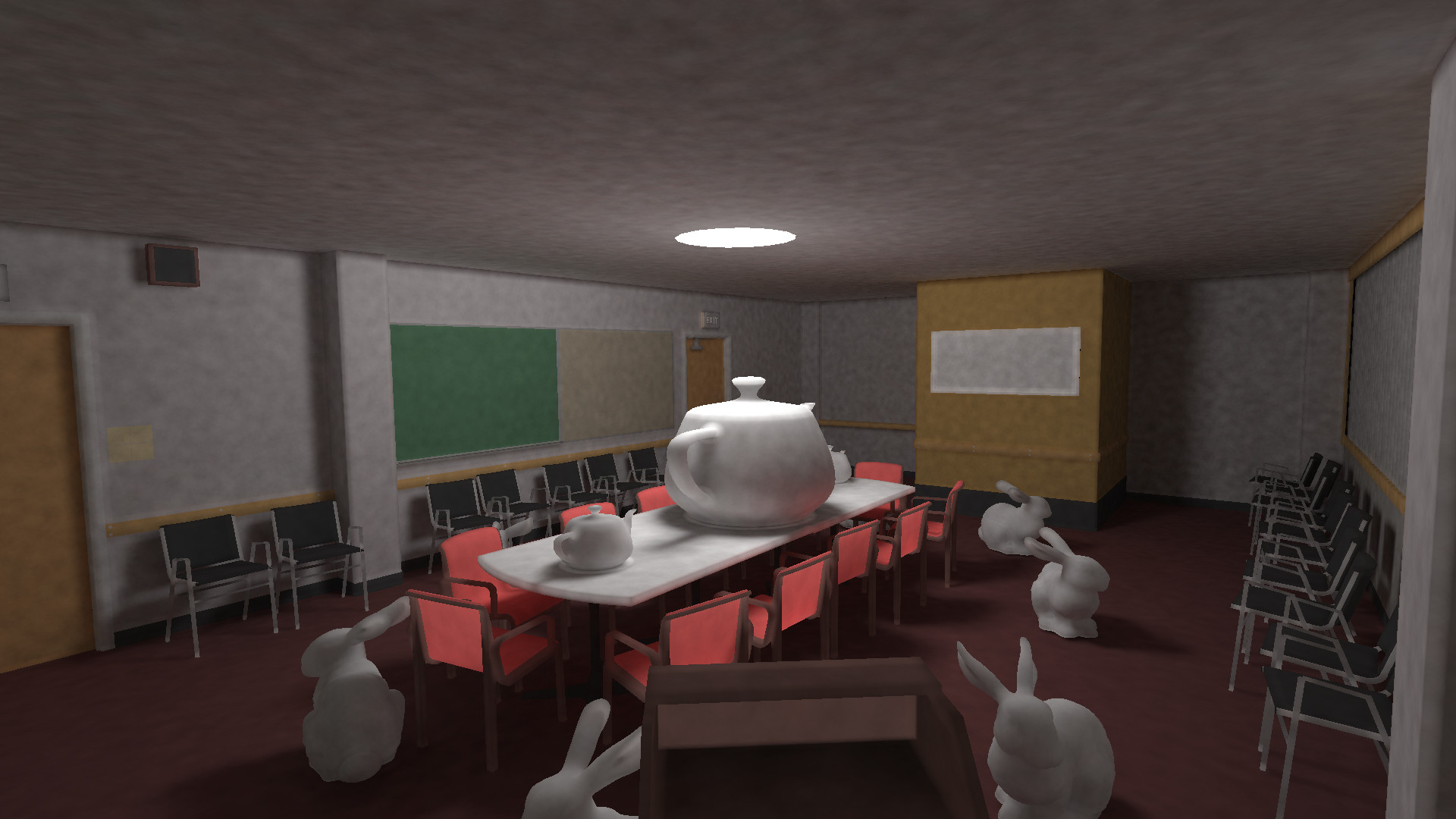}
  }
  \subfloat[Armadillo]{\label{fig:confMovingLecturer}
    \includegraphics[width=0.32\textwidth]{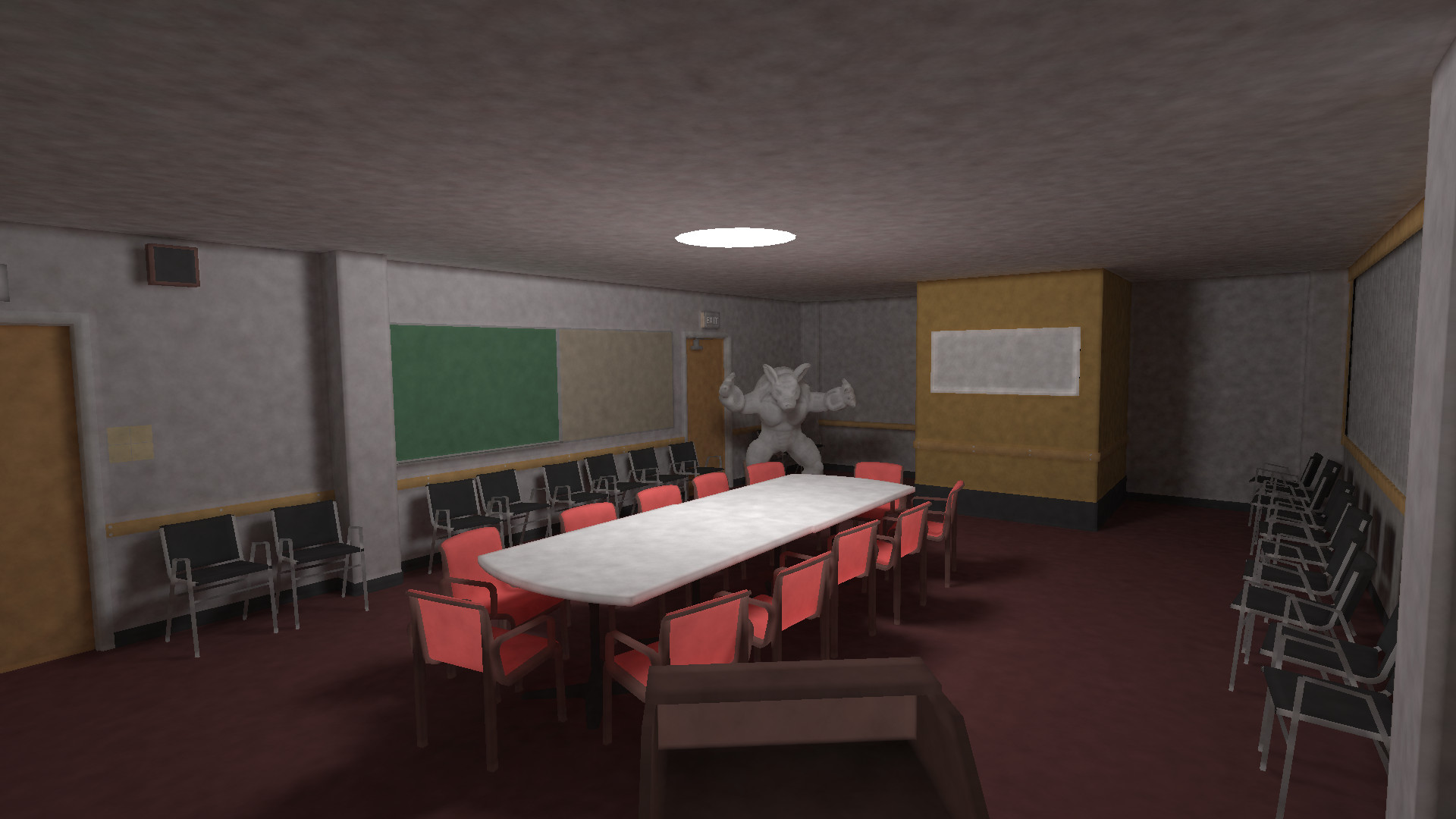}
  }
  \subfloat[Villa]{\label{fig:villa2MovingLight}
    \includegraphics[width=0.32\textwidth]{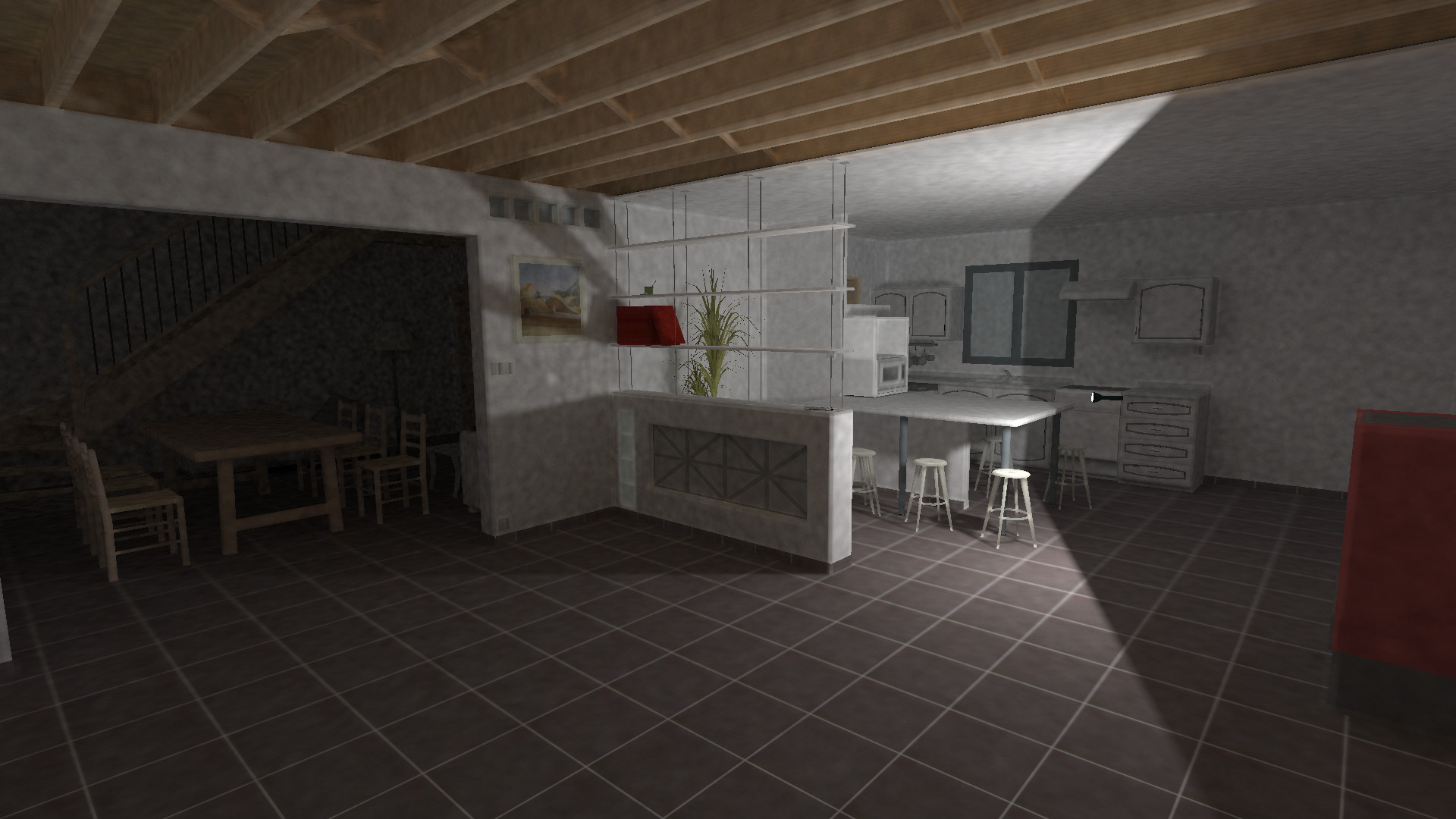}
  }
  \caption{Images of the scenes used in this paper.}
\end{figure*}

All presented results were rendered at a resolution of $1920 \times 1080$ on an
NVIDIA Titan X (Pascal architecture, 12 GB of VRAM). The tracing of the photons
was done using OptiX Prime 5.0.0~\cite{parker10}, whereas the path-reuse computations
were implemented using CUDA 9.1~\cite{Nickolls:2008:SPP:1365490.1365500}. We
compare our \enquote{naive} approach, presented in Section~\ref{sec:dynObjects},
to our error-based method, presented in Section~\ref{sec:errorBased} and to a
baseline, which consists in not reusing any information from previous frames and
re-tracing every single path each frame.

We tested our methods on different scenes:
\begin{description}
  \item[Merry-go-round] Conference, with a disc area light placed above the
    centre of the conference table, 3 scaling and rotating teapots placed on
    that table, around which 8 bunnies move as shown in
    Figure~\ref{fig:confMovingBunniesTeapots});
  \item[Armadillo] Conference scene with an armadillo moving from one door to
    the presenter stand, waiting there for a few seconds, then proceeding to
    the other door as shown in Figure~\ref{fig:confMovingLecturer}.
  \item[Villa] a small torchlight, made of a disc-shaped area light, is moving
    within the kitchen of a house, indirectly lighting the living room as shown
    in Figure~\ref{fig:villa2MovingLight}.
\end{description}

We recorded the first 30 seconds of the rendering of each scene, for the
baseline and our two methods; those videos can be found in the supplemental
materials. The configurations used (number of paths, resolution of the DM, etc.)
are the same as the ones mentioned in Figure~\ref{fig:graphs}.
Note that the time displayed in the top-right corner in the videos corresponds to the \emph{total frame time}, while Figure~\ref{fig:graphs} and \ref{fig:Breakdowns} both focus on only a few steps of the process, ignoring for example the time taken for splatting the photons ($\geq 130$~ms) as orthogonal to the reuse.

\subsection{Performance}

\begin{figure}
  \centering
  \subfloat[Our error-based method]{\label{fig:newBreakdown}
    \includegraphics[width=0.49\textwidth]{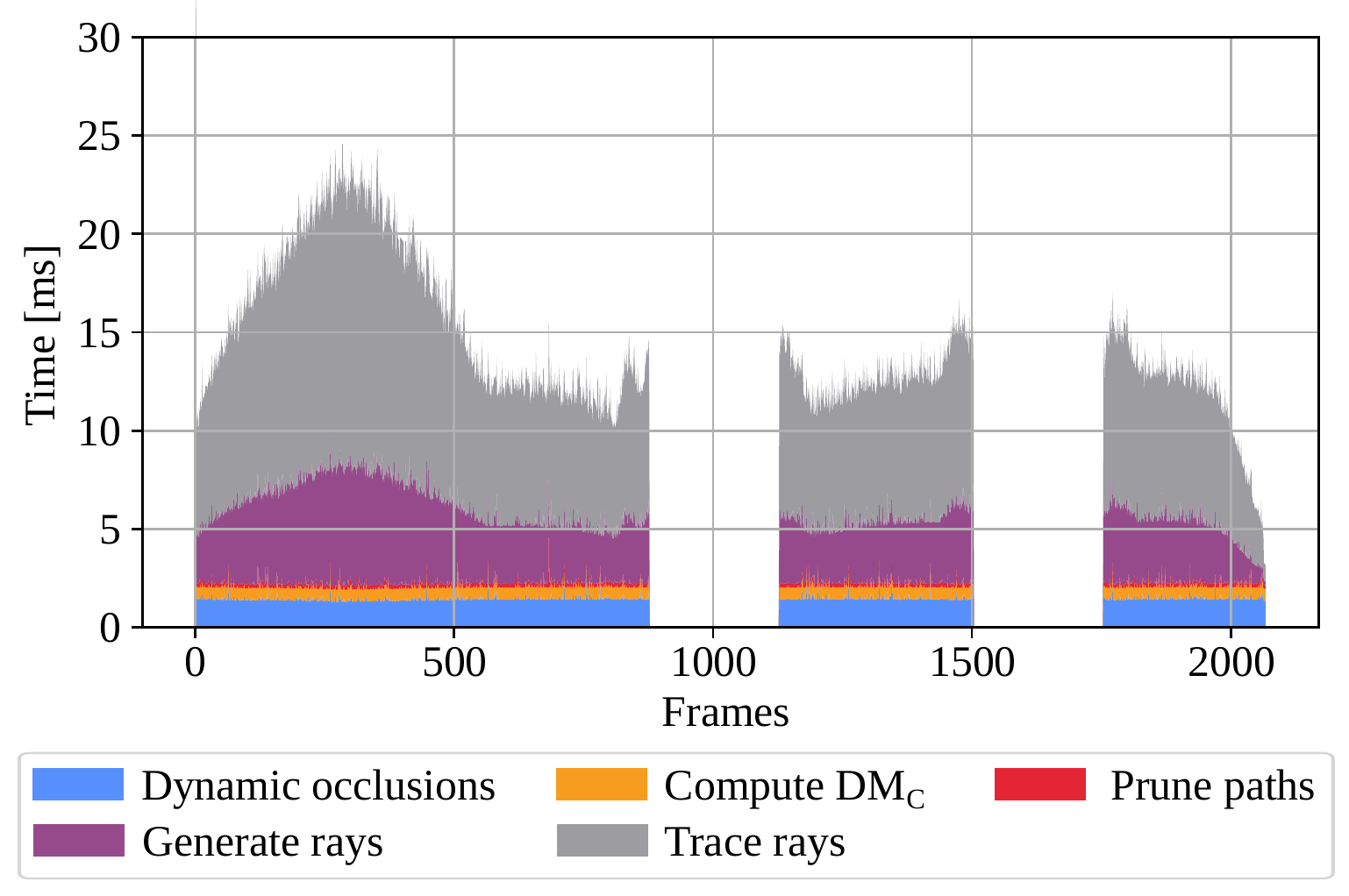}
  }

  \subfloat[Baseline]{\label{fig:refBreakdown}
    \includegraphics[width=0.49\textwidth]{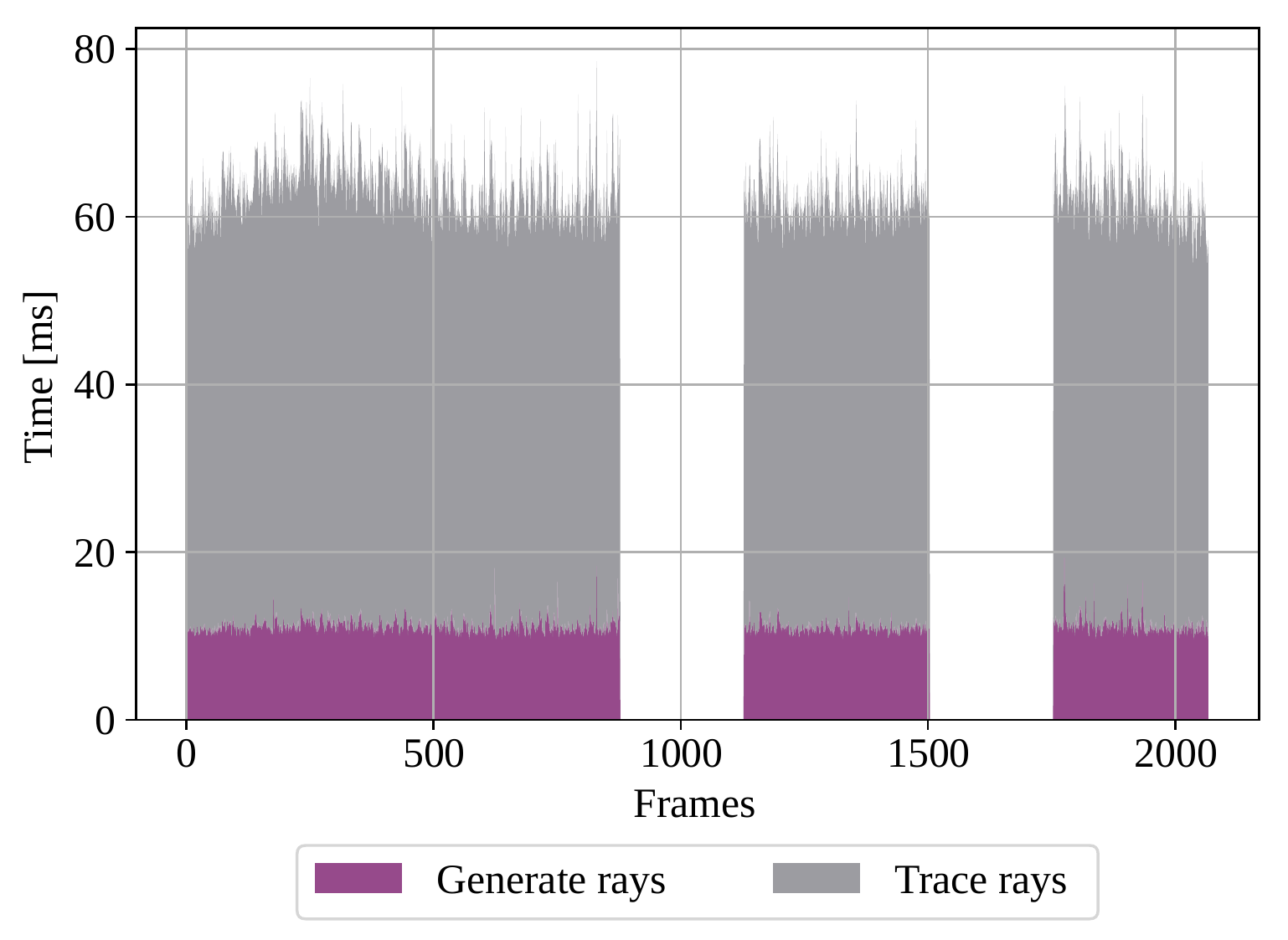}
  }
  \caption{Breakdowns of our method for reusing photons (top) and of the
    baseline (bottom), for the armadillo scene. In both cases, 2 m paths were
    traced with a maximum of 7 bounces, and for our method, the distribution
    map had a resolution of $8\times8\times64\times64$ while a 0.1\% error on
    the outgoing radiance was allowed on reused segments.
  }\label{fig:Breakdowns}
\end{figure}

The breakdowns presented in Figure~\ref{fig:Breakdowns} uses the different
categories presented in Figure~\ref{fig:algorithm_diagram}, but with the
modifications described in Section~\ref{sec:implementation}. So, for example,
the \enquote{update path origins} time is included within the \enquote{compute
$DM_C$} time, as they are implemented within the same kernel.

\begin{figure}
  \centering
  \subfloat[Merry-go-round]{\label{fig:tracingBunnies}
    \includegraphics[width=0.43\textwidth]{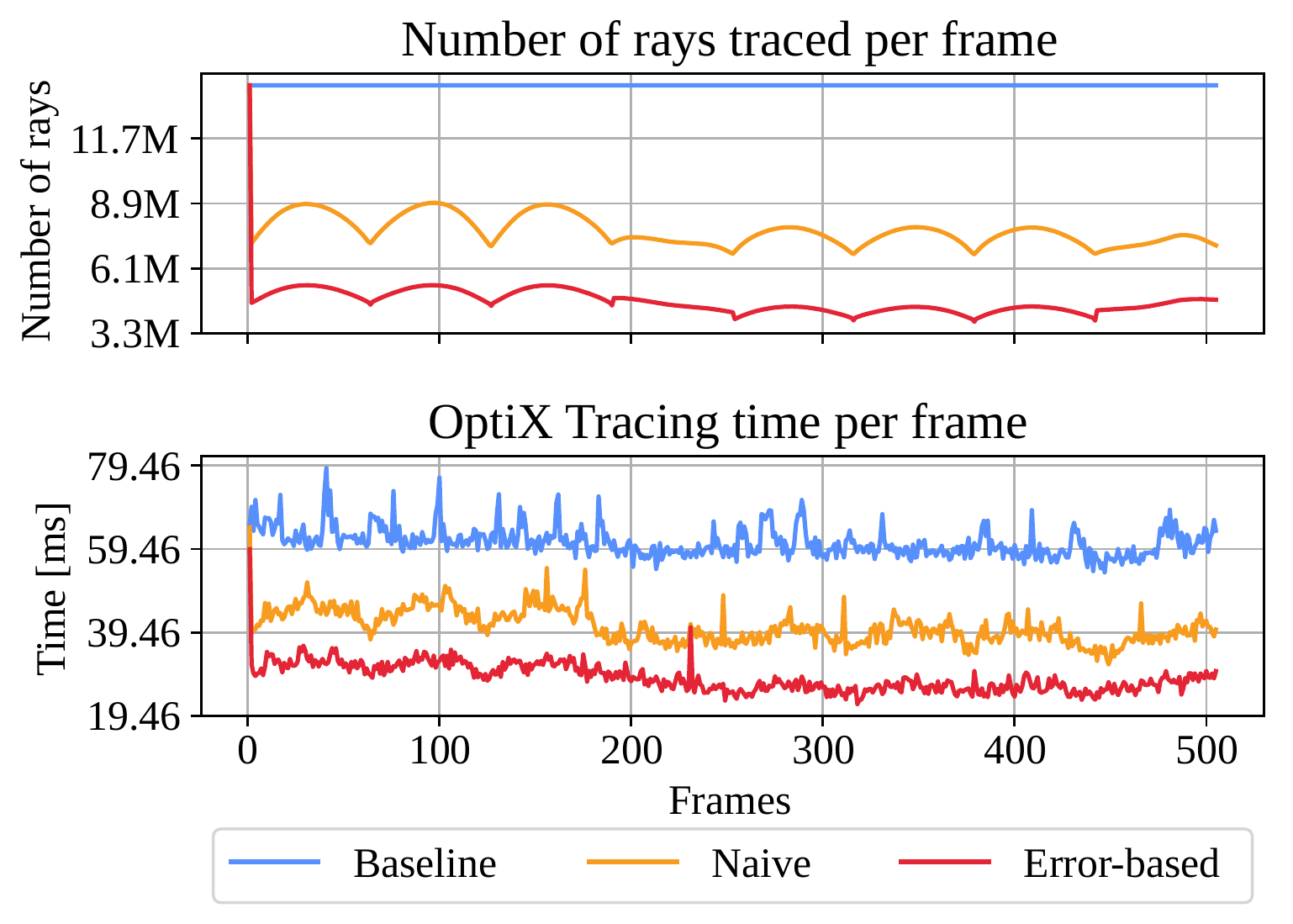}
  }

  \subfloat[Armadillo]{\label{fig:tracingLecturer}
    \includegraphics[width=0.43\textwidth]{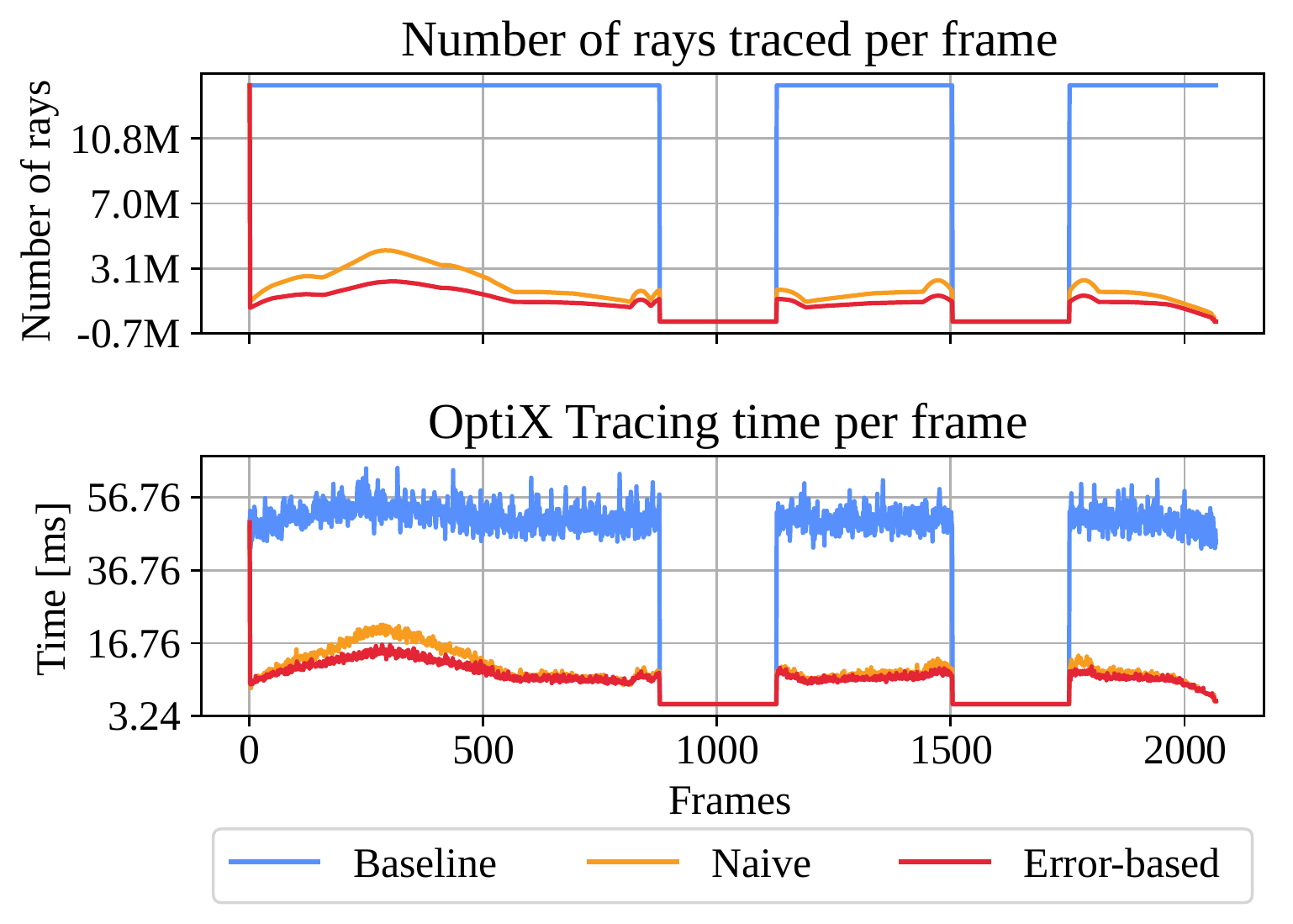}
  }

  \subfloat[Villa]{\label{fig:tracingVilla}
    \includegraphics[width=0.43\textwidth]{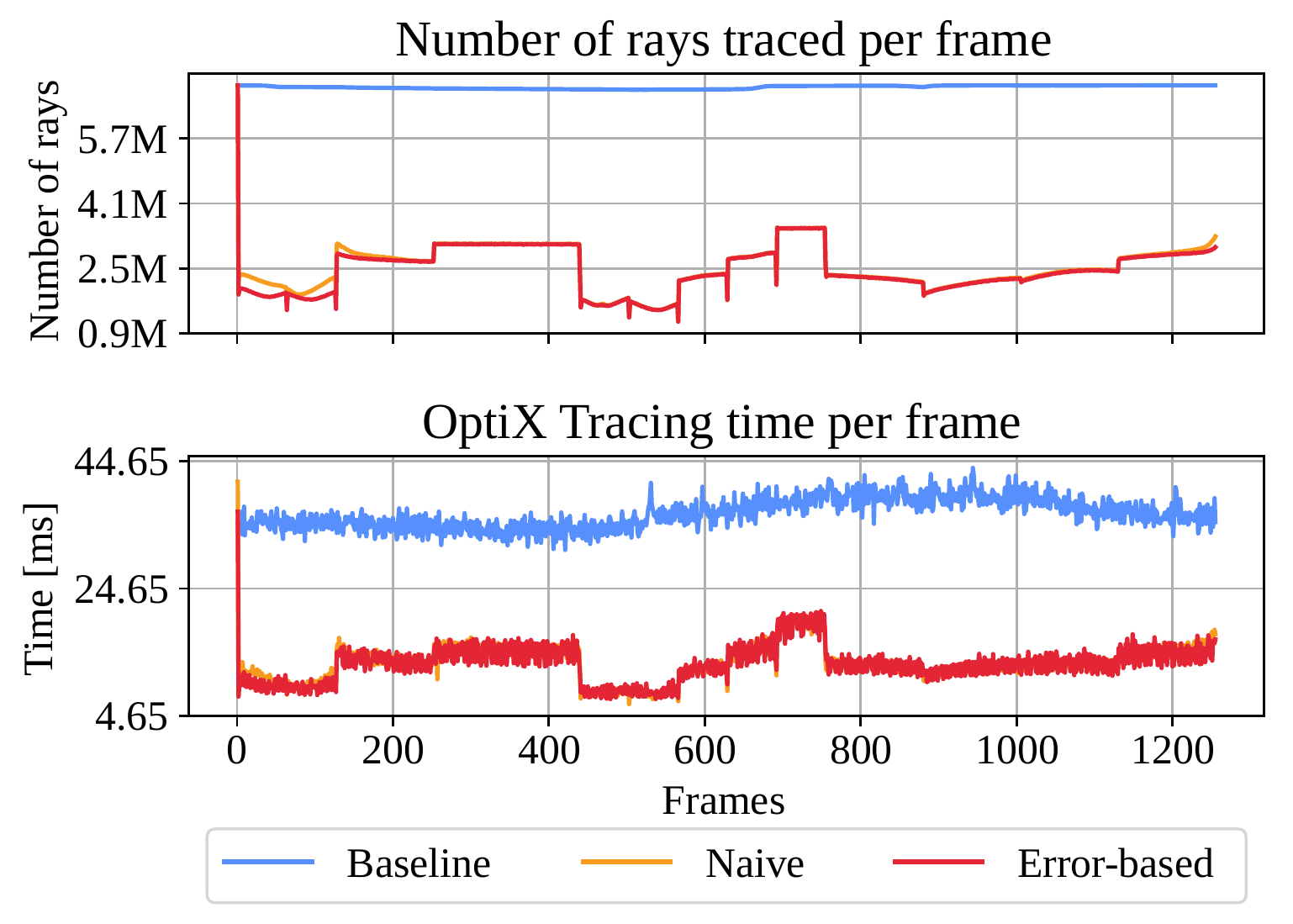}
  }
  \caption{Tracing time and number of rays compared to the baseline, for the
    different scenes. The merry-go-round and armadillo scenes both used 2
    million paths, whereas the villa has 1 million paths, but for all of them
    the paths contained at most 7 photons and the distribution map had a
    resolution of $8\times8\times64\times64$. For our error-based method, the
    energy threshold was set to 0.1\%.
  }\label{fig:graphs}
\end{figure}

Our two methods only differ in how they handle moving objects, but their
handling of moving lights is the same. This explains why there is no differences
between our two methods, neither in number of rays reused nor in tracing time,
in Figure~\ref{fig:tracingVilla}.

Even our naive method for dynamic objects already significantly reduces the
number of rays traced each frame, for example for the armadillo scene, it is
reduced by 5$\times$, as can be seen in Figure~\ref{fig:tracingLecturer}. This
does not translate into a 5$\times$ decrease in the time taken by OptiX prime
for tracing those rays, but into a 3$\times$ decrease instead. This could
come from more primary rays, proportionally, not being retraced, compared to
secondary rays, which are more expensive, as well as not taking special care to
maximise ray locality and coherency. Overall, our error-based method only
slightly improves the number of rays reused, except when the armadillo gets
close to the light source (around frame 250), where it retraces only half the
number of rays compared to our naive method.

The merry-go-round scene reduces the effectiveness of ray reuse, as many primary
rays will be
hitting a moving object, instantly invalidating the whole path. Despite that,
our naive method queries almost half as many rays as the baseline.
Furthermore, our error-based approach reuses close to 1.5$\times$ as many rays as
our naive approach, as seen in Figure~\ref{fig:tracingBunnies}.

Our different methods do add a small overhead compared to just re-tracing the paths
every frame. This overhead includes updating the path's origin, computing the
$DM_C$ and optimising it. On average the overhead is about 2.5~ms, compared to the average
baseline time of 60~ms, as shown in Figure~\ref{fig:Breakdowns}, and even
including this overhead our method still leads to an average 4$\times$ increase
in performance.

\subsection{Memory Consumption}

In this section we present the amount of memory being used for reusing photons from
previous frames. As reusing photons can be decoupled from the method used for
rendering using the photon map, we do not discuss the memory used for the rendering method.

\begin{table}
  \sisetup{
    table-number-alignment=right
  }
  \centering
  \caption{Memory-consumption (in \si{\mebi\byte}) breakdown when not reusing
    photons, reusing photons with moving lights and reusing photons with moving
    objects. In all scenarios, 5 millions paths containing each at most 7
    photons were considered; those paths were traced from a single disc-shaped
    area light, which was associated to a $32^4$ distribution map.
  }%
  \label{tab:memory}
  \begin{tabular}{l S S S}
    \toprule
    & {No reuse} & {Reuse lights} & {Reuse obj.} \\
    \midrule
    Path information   & {-}  & 19.07 & 19.07 \\
    Path origin pos. & {-}  & 57.22 & 57.22 \\
    Distribution maps  & {-}  & 8.000 & {-}   \\
    Pruned paths array & {-}  & 19.07 & {-}   \\
    \textbf{Sub-total} & {-} & 103.36 & 76.29 \\
    \midrule
    Photon map         & \multicolumn{3}{c}{1068} \\
    \textbf{Total}     & 1068 & 1171 & 1144   \\
    \bottomrule
  \end{tabular}
\end{table}

Path information is stored in a single 32-bit word, per path, as described in
Section~\ref{sec:implementation}. This compactness does introduce some
limitations, like being limited to at most 16 bounces, or to having at most
4 million cells in a distribution map, but those are not scenarios presented
in this paper and were done in order to improve performance and reduce memory
consumption. Those restrictions could be lifted by using more memory instead,
without needing to change the algorithm.

For each path, we also store the position on the light from which it was
emitted; this is only needed for area lights, as for point lights, it will
always be the same position as the light itself. One could avoid having to store
that information separately, by instead storing for each photon its incoming
direction, scaled by the distance between it and its predecessor, and its
position, allowing to recompute the origin point. However, this will make all
photons larger, resulting in an increased memory consumption.

A single $32^4$ DM is 4~\si{\mebi\byte}, but as each light gets two of them
(the current one and the expected one), the number reported is
8~\si{\mebi\byte}. Note that $DM_T$ could be compressed if memory
consumption is an issue, as, depending on the representation used, multiple
symmetries can be exploited. For example for a diffuse rectangular area light,
all points on its surface will have the same outgoing directions profile, so
only one set could be stored, bringing down the distribution map size from 4-D to
2-D. Also, if using an angular representation for the directions, the
values obtained for the partitioning along $\theta$ are the same for all $\phi$
partitions, bringing the dimensionality further down to 1. $DM_T$ can also be
computed as needed, to avoid having to store it.

When we need to process all pruned paths, i.e. paths that were marked during the
\enquote{prune paths} step (see Section~\ref{sec:dynLights}), we could go over
the path information attached to each path, and only process the ones marked.
However this could result in blocks with only a couple of active threads using
the GPU resources and preventing other blocks from running, whereas if combining
all active threads into as few blocks as possible, they could all run
simultaneously. So in order to achieve the latter, we maintain an array
containing all pruned paths, and process from the start only those paths, at the
cost of using more memory (a single 32-bit word per path).

In cases where paths do not bounce up to the limit, our photon map design
(described in Section~\ref{sec:implementation}) will be wasting some memory
space. It is however quite simple and allows straightforward accesses to any
photon of any path, and is quite efficient when processing all paths, at the
same $i$th bounce, simultaneously.

\section{Limitations}

\begin{description}
  \item[Glossy surfaces] If an intersection on a glossy surface is located on a
    static object and neither the incoming nor outgoing directions have changed,
    our method will be able to reuse those segments. However, if the above
    condition does not hold, then we might have to re-trace the outgoing ray,
    as even a small change in direction can lead to a large change
    in reflected energy.
  \item[Motion Blur] For this to be correct we would need to detect occlusions in between frames.
\end{description}

\section{Conclusion}

Path tracing for indirect illumination requires a substantial amount of
computation and in this paper we have shown how light transport paths can be reused
temporally by verifying the path segments.
In particular for moving lights we demonstrate that even though the light source
moves, we can still reuse photon paths coming from area light sources.
Furthermore when moving objects are present in the scene we demonstrate how
paths can be brute force tested against dynamic objects in a relatively short
amount of time compared to overall frame time.
By using an error threshold for path verification we further demonstrate that
path reuse can be improved and the number of retraced rays per frame can be
significantly reduced.
Path verification is particularly important for scenes with long paths where
reuse has an even greater impact on frame time.

Since our technique is focused on verifying the validity of paths, it would also
be applicable to camera paths for path tracing methods. For path tracing the
distribution map would be located on the near plane of the camera and the 2D
distribution map should behave similarly to that of a spotlight.

\begin{acks}
  In the villa scene, the \enquote{flash light}~\cite{modelFlashLight} model is
  courtesy of naves and the \enquote{maison à ossature
  bois}~\cite{modelMaisonOssature} model is courtesy of ADoc Envisioneer, both
  under CC Attribution 4.0. Conference, the bunny, the teapot and Armadillo are taken from
  Morgan McGuire’s Computer Graphics Archive~\shortcite{McGuire2017Data}.
  Pierre and Michael are sponsored by the \grantsponsor{VR2021005208}{Swedish Research Council}{https://www.vr.se/inenglish.4.12fff4451215cbd83e4800015152.html} under grant №~\grantnum{VR2021005208}{2014-5191}.
\end{acks}

\bibliographystyle{ACM-Reference-Format}
\bibliography{ms}

\end{document}